\documentclass[10pt,letterpaper]{article}
\usepackage{opex3} 

\usepackage{graphicx}
\newcommand{\VV}{\widetilde{V}}
\newcommand{\UU}{\widetilde{U}}
\newcommand{\x}{\mathbf{x}}
\newcommand{\be}{\begin{equation}}
\newcommand{\ee}{\end{equation}}
\newcommand{\bea}{\vspace{0.25cm}\begin{eqnarray}}
\newcommand{\eea}{\end{eqnarray}}
\begin{document}

\title{Analysis of the possibility of analog detectors calibration by exploiting Stimulated Parametric Down
Conversion }

\author{Giorgio Brida\textsuperscript{1},
Maria Chekhova\textsuperscript{2}, Marco
Genovese\textsuperscript{1},\\
 Ivano Ruo-Berchera\textsuperscript{1}}
\address{$^1$Istituto Nazionale di Ricerca Metrologica, Strada delle Cacce 91, 10135 Torino, Italy}
\address{$^2$Physics Department, M.V. Lomonosov State University, 119992
Moscow, Russia. }
\email{i.ruoberchera@inrim.it}

\begin{abstract}
Spontaneous parametric down conversion (SPDC) has been largely
exploited as a tool for absolute calibration of photon-counting
detectors, i.e detectors registering very small photon fluxes. In
\cite{systematic} we derived a method for absolute calibration of
analog detectors using SPDC emission at higher photon fluxes,
where the beam is seen as a continuum by the detector.
Nevertheless intrinsic limitations appear when high-gain regime of
SPDC is required to reach even larger photon fluxes. Here we show
that stimulated parametric down conversion allow one to avoid this
limitation, since stimulated photon fluxes are increased by the
presence of the seed beam.
\end{abstract}



\section{Introduction}

Prompted by the necessity of a precise calibration of
photo-detectors both in the analog and photon-counting regime
\cite{ba,ge,QC}, we presented recently \cite{systematic} a
detailed theoretical analysis on the possibility to use the
correlations of spontaneous parametric down conversion (SPDC)
light \cite{bp1,Burnham} for calibrating analog detectors, as an
extension of the technique developed in photon counting regime
\cite{klysh,malygin,KP,alan,Brida1,Ginzburg,poc,Brida3,Brida4}.
However, while the method is suitable for analog calibration at
relatively low gain, which means a photon flux lower than
$10^{10}$ photon/s, in higher-gain regime it is limited by the
difficulty of collecting the same correlated modes in the two
branches \cite{systematic}.

In this paper we present a detailed theoretical analysis of a
scheme based on the stimulated PDC (on the other hand a discussion
of the uncertainty budget is left to a specifically addressed
paper \cite{JMO}). We show that this scheme allow to overcome the
problem mentioned above. Indeed, in this case the photon fluxes
can be varied by varying the power of the coherent seed beam,
without increasing the parametric gain $G$. On the other hand, the
SPDC non-classical correlation at single-photon level which
enables the absolute calibration of counting detectors survives,
in some form, when a coherent seed is injected and the photon flux
becomes macroscopic.

\begin{figure}[tbp]
\par
\begin{center}
\includegraphics[angle=0, width=9cm, height=6 cm]{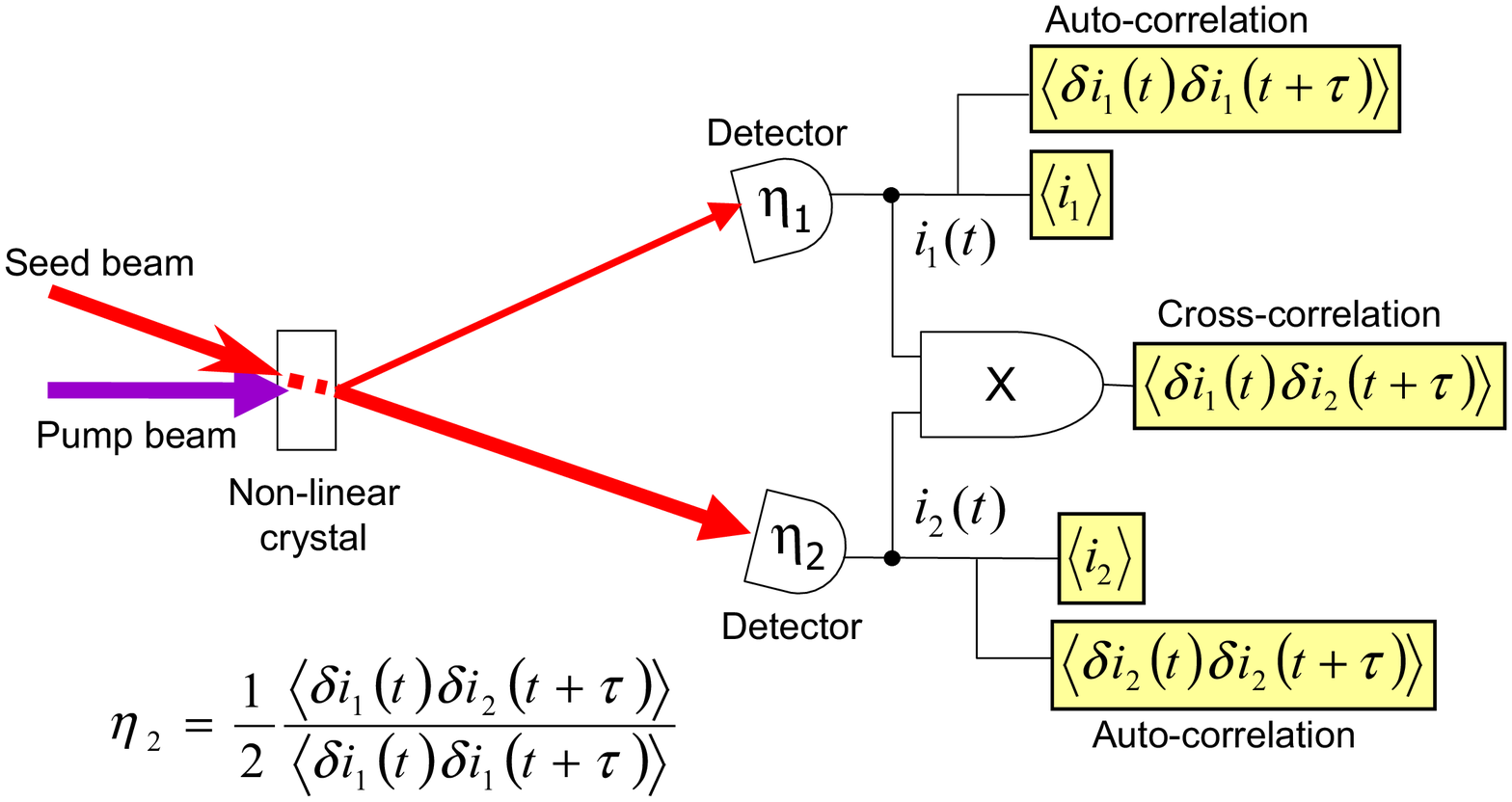}
\caption{\textsf{Scheme for absolute calibration of analog
detectors by using stimulated PDC. The quantum efficiency of
detector 2, collecting the seed beam and the stimulated emission,
is estimated by the ratio between the cross-correlation function
of the photocurrents  and the auto-correlation of the photocurrent
$i_{1}$.}} \label{calib_stim}
\end{center}
\end{figure}

\section{Analog detection and quantum efficiency}

The schematic set-up for calibration of photodetectors by using
PDC is shown in Fig. 1. We model the photodetection process in the
analog regime as a random pulse train \cite{Soda}

\begin{eqnarray}
  i(t)=\sum_{n}q_{n}f(t-t_{n})\nonumber,
\end{eqnarray}
i.e. a very large number of discrete events at random times of
occurrence $t_{n}$. The pulse shape $f(t)$ is determined by the
transit time of charge carriers. We assume that $f(t)$ is a fixed
function with characteristic width $\tau_{p}$ and a unit area.
 The pulse amplitude $q_{n}$ is a random variable in
order to account for a possible current gain by avalanche
multiplication. The statistical nature of the multiplication
process gives an additional contribution to the current
fluctuations \cite{MNUAD}. In an ideal instantaneous photocell
response, without avalanche gain, all values $q_{n}$ are equal to
the charge $e$ of a single electron and $f(t)\sim \delta(t)$. In
the case of ideal quantum efficiency, since the probability
density of observing a photon at time $t$ at detector $D_{j}$
($j=1,2$) is related to the quantum mean value $\langle
F_{j}(t)\rangle$ of the photon flux, we calculate the average
current output of $Dj$ as
\begin{eqnarray}\label{curr}
\langle i_{j}\rangle =\sum_{n}\langle q_{jn}f(t-t_{n})\rangle
=\int dt_{n}\langle q_{j}\rangle f(t-t_{n})\langle
\widehat{F}_{j}(t_{n})\rangle\nonumber\\
\end{eqnarray}
where the factor $\langle q_{j}\rangle$ is the average charge
produced in a detection event. We have assumed the response
function for the two detectors to be the same,
$f_{1}(t)=f_{2}(t)=f(t)$.

At the same time, the auto-correlation and the cross-correlation
functions for the currents can be expressed as

\begin{eqnarray}\label{i_{j}i_{k}}
\langle i_{j}(t)i_{k}(t+\tau)\rangle =\sum_{n,m}\langle
q_{jn}q_{km}
f(t-t_{n})f(t-t_{m}+\tau)\rangle\nonumber\\
=\int\int dt_{n}dt_{m} \langle q_{j} q_{k}\rangle
f(t-t_{n})f(t-t_{m}+\tau)\langle
\widehat{F}_{j}(t_{n})\widehat{F}_{k}(t_{m})\rangle,\nonumber\\
\end{eqnarray}
respectively for $j=k$ where $\langle
\widehat{F}_{j}(t_{n})\widehat{F}_{j}(t_{m})\rangle$ is the
auto-correlation function of the photon flux at detector $j$, and
for $j\neq k$ where $\langle
\widehat{F}_{j}(t_{n})\widehat{F}_{k}(t_{m})\rangle$ is the
cross-correlation between the fluxes incident on the two different
detectors. By the substitution $\widehat{F}_{j}\equiv\langle
\widehat{F}_{j}\rangle+\delta\widehat{F}_{j}$, it is convenient to
express them as

\begin{eqnarray}\label{F_j(tn)F_k(tm)}
\langle \widehat{F}_{j}(t_{n})
\widehat{F}_{k}(t_{m})\rangle&=&\langle
\widehat{F}_{j}\rangle\langle \widehat{F}_{k}\rangle+\langle
\widehat{F}_{j}\rangle
\delta(t_{n}-t_{m})\delta_{jk}\nonumber\\
&+&\langle:\delta\widehat{F}_{j}(t_{n})
\delta\widehat{F}_{k}(t_{m}):\rangle.
\end{eqnarray}
The second contribution, proportional to the photon flux when
$j=k$, represents the the intrinsic and unavoidable component of
the fluctuation that does not depend on the specific property of
the field since it generates from the commutation relation of the
quantum fields in the free space. The third one is the normally
ordered correlation function of the fluctuation. Actually it is a
function just of the difference $t_{n}-t_{m}$ and its typical
variation scale provides the coherence time $\tau_{coh}$ of the
PDC radiation.

Now we introduce the quantum efficiency $\eta_{j}$ of detector
$D_{j}$, defined as the number of pulses generated per incident
photon. In \cite{systematic} a real detector is modelled, as
usual, with an ideal one ($\eta=1$) preceded by a beam splitter
with the intensity transmission coefficient equal to the quantum
efficiency of the real detector \cite{photon-noise}. Within this
picture it is possible to take into account the quantum efficiency
by the following substitutions:
\begin{eqnarray}\label{eta}
\langle \widehat{F}_{j}(t)\rangle &\longrightarrow&\eta_{j}\langle
\widehat{F}_{j}(t)\rangle
\nonumber\\
\langle:\widehat{F}_{j}(t)\widehat{F}_{k}(t'):\rangle&\longrightarrow&\eta_{j}\,\eta_{k}\langle
:\widehat{F}_{j}(t)\widehat{F}_{k}(t'):\rangle.
\end{eqnarray}

Thus, $\langle \widehat{F}_{j}(t)\rangle$ being  time independent,
according to Eq. (\ref{curr}) we obtain:
\begin{equation}\label{curr-time}
  \langle i_{j}\rangle=\eta_{j}\langle q_{j}\rangle\langle
F_{j}\rangle.
\end{equation}

Eq. (\ref{i_{j}i_{k}}) becomes

\begin{eqnarray}\label{i_{j}i_{k}new}
\langle i_{j}(t)i_{k}(t+\tau)\rangle =\langle i_{j}\rangle\langle
i_{k}\rangle+\eta_{j}\langle q_{j}^{2}\rangle\mathcal{F}(\tau)
\langle F_{j}\rangle
\delta_{jk}\nonumber\\
+\eta_{j}\,\eta_{k}\langle q_{j}q_{k}\rangle\int\int
dt_{n}dt_{m}\nonumber\\
f(t-t_{n})f(t-t_{m}+\tau)\langle:\delta\widehat{F}_{j}(t_{n})\delta\widehat{F}_{k}(t_{m}):\rangle,\nonumber\\
\end{eqnarray}
where we introduced the convolution of the response function of
detectors $\mathcal{F}(\tau)= \int dt f(t)f(t+\tau)$.

\section{Correlation functions of stimulated PDC}

Here we consider a type I PDC process. In the limit of
monochromatic and plane-wave pump approximation, only pairs of
modes with opposite transverse wave vectors, $\mathbf{q}$ and
$-\mathbf{q}$, and with conjugate frequencies,
$\omega_{1}=\omega_{pump}/2-\Omega$ and
$\omega_{2}=\omega_{pump}/2+\Omega$, are coupled such that the
energy and momentum conservation hold. The equations describing
the down conversion process are the input-output transformations
\begin{eqnarray}\label{input-output}
\widehat{a}_{1}^{out}(\mathbf{q},\Omega)=
U_{1}(\mathbf{q},\Omega)\;\widehat{a}_{1}^{in}(\mathbf{q},\Omega)+
V_{1}(\mathbf{q},\Omega)\;\widehat{a}_{2}^{in\dag}(-\mathbf{q},-\Omega)\nonumber\\
\widehat{a}_{2}^{out}(\mathbf{q},\Omega)=
U_{2}(\mathbf{q},\Omega)\;\widehat{a}_{2}^{in}(\mathbf{q},\Omega)+
V_{2}(\mathbf{q},\Omega)\;\widehat{a}_{1}^{in\dag}(-\mathbf{q},-\Omega)\nonumber\\
\end{eqnarray}
linking the fields $\widehat{a}_{1}^{out}$ and
$\widehat{a}_{2}^{out}$ after the non-linear interaction to the
fields $\widehat{a}_{1}^{in}$ and $\widehat{a}_{2}^{in}$ before
the interaction started. The coefficients
$U_{k}(\mathbf{q},\Omega)$ and $V_{k}(\mathbf{q},\Omega)$ must
satisfy the properties

\begin{eqnarray}\label{unitary cond}
|U_{k}(\mathbf{q},\Omega)|^{2}-|V_{k}(\mathbf{q},\Omega)|^{2} =  1, \quad (k=1,2)  \nonumber  \\
U_{1}(\mathbf{q},\Omega)V_{2}(-\mathbf{q},-\Omega)=U_{2}(-\mathbf{q},-\Omega)V_{1}(\mathbf{q},\Omega),
\end{eqnarray}
which guarantee the conservation of the free-field commutation
relations
$$[\widehat{a}_{j}^{out}(\mathbf{q},\Omega),\widehat{a}_{k}^{out\dag}(\mathbf{q}',\Omega')]=
\delta_{j,k}\delta_{\mathbf{q},\mathbf{q}'}\delta_{\Omega,\Omega'}$$
$$[\widehat{a}_{j}^{out}(\mathbf{q},\Omega),\widehat{a}_{k}^{out}(\mathbf{q'},\Omega')]=0.$$
Here, $U_{k}$ and $V_{k}$ define the strength of the process and
at the same time the range of transverse momentum and frequencies
in which it takes place and thus are named gain functions. In
fact, the finite length of the crystal introduces a partial
relaxation of phase matching condition concerning the third
component of the momenta. By selecting a certain frequency, the
transverse momentum uncertainty, i.e. the angular dispersion of
the emission direction, is not null ($\Delta
q\sim(l\tan\overline{\vartheta})^{-1}$ for the non-collinear PDC,
where $l$ is the crystal length and $\overline{\vartheta}$ is the
central angle of propagation with respect to the pump direction).
On the contrary, by fixing the transverse momentum $\mathbf{q}$,
or equivalently a direction of propagation $\vartheta$, the
spectral bandwidth $\Delta\Omega$ is proportional to $l^{-1/2}$,
and typically $1/\Delta\Omega\sim10^{-13}$s for type I.

Hereinafter we will focus on the properties of the far field,
observed at the focal plane of a thin lens of focal length $f$,
placed at distance $f$ from the crystal. The  spatial distribution
of the far field is, in this case, the Fourier transform of the
field just at the output face of the crystal. This special imaging
configuration is convenient to show the basic of calculation,
nonetheless the validity of the results is more general. Thus, any
transverse mode $\mathbf{q}$ is associated with a single point
$\mathbf{x}$ in the detection (focal) plane according to the
geometric transformation $\mathbf{q}\rightarrow \
2\pi\mathbf{x}/(\lambda f)$. The far field operator in the
space-temporal domain is therefore
\begin{equation}\label{stim0}
\widehat{B}_{k}(\mathbf{x},t)=\sum_{\Omega}
e^{-i\;\Omega\;t}\;\widehat{a}_{k}^{out}\left(\frac{2\pi}{\lambda
f}\mathbf{x},\Omega\right)\;.
\end{equation}

The mean value of the operator $\widehat{I}_{k}(\x,t)\equiv
\widehat{B}^{\dag}_{k}(\mathbf{x},t)
\widehat{B}_{k}(\mathbf{x},t)$ is

\begin{eqnarray}\label{stim1}
\langle \widehat{I}_{k}(\x,t)\rangle&=&\sum_{\Omega\,\Omega'}
e^{-i(\Omega'-\Omega)\;t}\nonumber\\
&&\left\langle\widehat{a}_{k}^{out\dag}\left(\frac{2\pi}{\lambda
f}\mathbf{x},\Omega\right)
\widehat{a}_{k}^{out}\left(\frac{2\pi}{\lambda
f}\mathbf{x},\Omega'\right)\right\rangle\;,
\end{eqnarray}
representing the intensity profile of the emission in the
detection plane. This mean value has to be calculated over the
following initial state, in which the field 1 is in the vacuum
state whereas the field 2 is in a multi-mode coherent state:
\begin{equation}\label{stim2}
\left|\psi_{in}\right\rangle=
\left|0\right\rangle_{1}\bigotimes_{\mathbf{q},\Omega}\left|\alpha(\mathbf{q},\Omega)\right\rangle_{2},
\end{equation}
where
$$\alpha(\mathbf{q},\Omega)=\alpha(\mathbf{q})\,\delta_{\Omega_{0}\Omega}$$
is the complex parameter associated to this state. The seed beam
is therefore represented as a coherent state of field 2 with fixed
frequency $\Omega_{0}$ and a certain distribution of the
transverse momentum $\mathbf{q}$. Here we assume that, for the
modes that are stimulated by the seed, the spontaneous component
of the emission is negligible with respect to the stimulated one.
This corresponds to the assumption
$|V_{k}(\mathbf{q},\Omega_{0})|^{2}\ll1$ and, at the same time,
$|\alpha(\mathbf{q})|^{2}\gg1$, that is a typical experimental
situation in which a few millimeters length crystal is pumped by a
continuous pump and the seed has a power just around the microwatt
or more. The intensities (\ref{stim1}) of the two stimulated beams
after the crystal can be evaluated by using Eq.s
(\ref{input-output}), with the substitution $\mathbf{q}\rightarrow
2\pi\mathbf{x}/(\lambda f)$. This leads to

\begin{equation}\label{stim3}
\langle \widehat{I}_{1}(\x,t)\rangle\approx
|\VV_{1}(\mathbf{x},-\Omega_{0})|^{2}
|\widetilde{\alpha}(-\mathbf{x})|^{2}
\end{equation}
\begin{eqnarray}\label{stim4}
\langle
\widehat{I}_{2}(\x,t)\rangle&\approx&|\UU_{2}(\mathbf{x},\Omega_{0})|^{2}
|\widetilde{\alpha}(\mathbf{x})|^{2}\nonumber\\
&=&\left(|\VV_{2}(\mathbf{x},\Omega_{0})|^{2}+1\right)
|\widetilde{\alpha}(\mathbf{x})|^{2}
\end{eqnarray}
where we defined
\begin{eqnarray}
\widetilde{U}_{k}(\x,\Omega)=U_{k}\left(\frac{2\pi}{\lambda f}\x,\Omega\right),\nonumber\\
\widetilde{V}_{k}(\x,\Omega)=V_{k}\left(\frac{2\pi}{\lambda f}\x,\Omega\right),\nonumber\\
\widetilde{\alpha}(\mathbf{x})=\alpha\left(\frac{2\pi}{\lambda
f}\mathbf{x}\right).
\end{eqnarray}

We used the first property of the gain functions in (\ref{unitary
cond}) into the last line of Eq. (\ref{stim4}), in order to show
explicitly the two contributions to the beam 2, one given by the
original coherent seed and the other from the stimulated emission,
proportional to the parametric gain $G\equiv
max|V_{k}(\mathbf{q},\Omega_{0})|^{2}$. We notice that, aside from
the spontaneous emission that is neglected, the generated beams
conserve the original momentum distribution of the seed but
weighted according to the gain function.

The normal-ordered correlation function of the intensities
fluctuation is defined as

\begin{eqnarray}\label{dI(x,t)dI(x',t')stim}
\langle: \delta \widehat{I}_{k}(\x,t) \delta
\widehat{I}_{j}(\x',t'):\rangle&=&\langle: \widehat{I}_{k}(\x,t)
\widehat{I}_{j}(\x',t'):\rangle-\nonumber\\
&-&\langle \widehat{I}_{k}(\x,t)\rangle\langle
\widehat{I}_{j}(\x',t')\rangle
\end{eqnarray}
The calculation leads to the following normal-ordered mean values:
\begin{equation}\label{stim5}
\langle
:\delta\widehat{I}_{1}\,(\x,t)\delta\widehat{I}_{1}(\x',t'):\rangle\approx0\,,
\end{equation}
\begin{eqnarray}\label{stim6}
\langle
:\delta\widehat{I}_{2}(\x,t)\;\delta\widehat{I}_{2}(\x',t'):\rangle&\approx&\,\sum_{\Omega}e^{-i(\Omega_{0}-\Omega)\;(t-t')}
 |\widetilde{\alpha}(\mathbf{x})|^{2}\nonumber\\
&&|\VV_{2}(\mathbf{x},\Omega)|^{2}\;|\UU_{2}(\mathbf{x},\Omega_{0})|^{2}\delta_{\x\x'}\nonumber\\
&&+\;c.c.
\end{eqnarray}
Fluctuations of the stimulated field 1 are negligible, because
they have the same order of magnitude of the spontaneous emission.
Concerning the cross-correlation we have:
\begin{eqnarray}\label{stim7}
\langle
:\delta\widehat{I}_{1}(\x,t)\;\delta\widehat{I}_{2}(\x',t'):\rangle&\approx&\,\sum_{\Omega}e^{-i(\Omega_{0}+\Omega)\;(t-t')}
 |\widetilde{\alpha}(-\mathbf{x})|^{2}\nonumber\\
 &&\UU_{2}(-\mathbf{x},\Omega_{0})\;\UU^{\star}_{2}(-\mathbf{x},-\Omega)\nonumber\\
&&\VV_{1}(\mathbf{x},-\Omega_{0})\;\VV^{\star}_{1}(\mathbf{x},\Omega)\delta_{-\x\x'}\nonumber\\
&&+\;c.c.
\end{eqnarray}
Let us consider two detectors, $D_{1}$ and $D_{2}$, to register
the photons crossing the regions $R_{1}$ and $R_{2}$ centered on
the two correlated bright beams in the detection plane. Let the
two areas be much larger than the dimension of the seed beam on
the detection plane. This simply means to collect all the bright
components of the emission, corresponding to the modes stimulated
by the seed.

We move now from the discrete modes representation of the
electromagnetic field, in which it is artificially enclosed in a
fictitious cube of side $L$, to a continuous set of modes for
$L\rightarrow\infty$. Thus, in the following, the $\Omega$-sum and
the Kronecker function $\delta_{\x\x'}$ in Eq.s (\ref{stim6}) and
(\ref{stim7}) are substituted respectively with the integral over
$\Omega$ and the Dirac function $\delta(\x-\x')$. The fact that
the second-order correlation function has an unphysical spatial
behaviour, dominated by a delta function, comes from the starting
assumption of a plane-wave pump, having infinite transverse
dimension. In a realistic case the delta function is replaced by
the Fourier transform of the pump transverse profile.

The quantum mean values of the photon fluxes
$F_{j}\equiv\int_{R_{j}}\widehat{I}_{j}(\x,t)d\x $ reaching
detectors 1 and 2 can be obtained integrating equations
(\ref{stim3}) and (\ref{stim4}) respectively on $R_{1}$ and
$R_{2}$. By applying relations (\ref{unitary cond}) we have

\begin{equation}\label{stim8}
\langle
F_{1}\rangle\approx\int_{R_{1}}d\mathbf{x}\:|\VV_{1}(\mathbf{x},-\Omega_{0})|^{2}
|\widetilde{\alpha}(-\mathbf{x})|^{2}\,,
\end{equation}
\begin{equation}\label{stim9}
 \langle F_{2}\rangle\approx\int_{R_{2}}d\mathbf{x}\:\left(|\VV_{2}(\mathbf{x},\Omega_{0})|^{2}+1\right)
|\widetilde{\alpha}(\mathbf{x})|^{2}\,.
\end{equation}
At the same time integrating Eq.s (\ref{stim5}) and (\ref{stim6})
over $R_{1}\times R_{1}$ and $R_{2}\times R_{2}$ respectively, we
have the following auto-correlation functions for the photon
fluxes

\begin{eqnarray}\label{stim10}
\langle :\delta F_{1}(t) \delta F_{1}(t'):\rangle\approx 0
\end{eqnarray}
\begin{eqnarray}\label{stim11}
\langle: \delta F_{2}(t) \delta
F_{2}(t'):\rangle&\approx&\sum_{\Omega}e^{-i(\Omega_{0}-\Omega)\;(t-t')}\nonumber\\
&&\int_{R_{2}}d\mathbf{x}\:|\VV_{2}(\mathbf{x},\Omega)|^{2}
|\widetilde{\alpha}(\mathbf{x})|^{2}+c.c.\,,\nonumber\\
\end{eqnarray}
 Here and in the following, we have assumed that in the limit of small $G$ the gain
functions $|\UU_{j}(\mathbf{x},\Omega)|\sim1$, according to the
relations (\ref{unitary cond}). Then, integrating Eq.
(\ref{stim7}) over $R_{1}\times R_{2}$ gives the following
cross-correlation:
\begin{eqnarray}\label{stim12}
\langle: \delta F_{1}(t) \delta F_{2}(t'):
\rangle\approx\sum_{\Omega}e^{-i(\Omega_{0}+\Omega)\;(t-t')}
\int_{R_{1}}d\mathbf{x}\nonumber\\
|\widetilde{\alpha}(-\mathbf{x})|^{2}\VV_{1}(\mathbf{x},-\Omega_{0})\;\VV^{\star}_{1}(\mathbf{x},\Omega)+c.c.\,.\nonumber\\
\end{eqnarray}
Although the seed beam is monochromatic with frequency
$\Omega_{0}$, the temporal width of both auto-correlation and
cross-correlation in a fixed point $\x$ still depends on the
spectral bandwidth $\Delta\Omega$ of the gain functions. This
explains why they originate from the correlation between the
monochromatic seed and the spontaneous emission that is broadband.
Therefore, the coherence time of the stimulated emission is still
of the order of $\tau_{coh}=1/\Delta\Omega$, analogously to the
spontaneous case.

\section{Quantum efficiency estimation}

The auto-correlation of the current fluctuations $\delta
i_{1}(t)=i_{1}(t)-\langle i_{1}\rangle$ can be calculated by
introducing the result in Eq. (\ref{stim10}) in
(\ref{i_{j}i_{k}new}), obtaining

\begin{eqnarray}\label{stimul-self-corr}
\langle \delta i_{1}(t) \delta
i_{1}(t+\tau)\rangle=\eta_{1}\langle q_{1}^{2}\rangle
\mathcal{F}(\tau) \langle F_{1}\rangle.
\end{eqnarray}
Under the condition of small gain ($G\ll1$) and large intensity of
the seed ($|\alpha|^{2}\gg 1$) the fluctuation of the current at
detector $D_{1}$ is dominated by the shot noise component.

By substituting Eq (\ref{stim12}) in (\ref{i_{j}i_{k}new}), under
the same condition, for $j=1$ and $k=2$, one has
\begin{eqnarray}\label{stimul-cross-corr}
\langle \delta i_{1}(t) \delta i_{2}(t+\tau)\rangle=2
\eta_{1}\eta_{2}\langle q_{1}\rangle\langle q_{2}\rangle
\mathcal{F}(\tau)\langle F_{1}\rangle
\end{eqnarray}
where we assumed $\tau_{p}\gg\tau_{coh}$. We stress that it is the
usual situation, the coherence time of SPDC being on the order of
picoseconds or less and the typical resolving time of detectors on
the order of nanosecond. In this case any fluctuations in the
light power are averaged over $\tau_{p}$.

The cross-correlation function of the fluctuations has the same
form as the one obtained in the case of spontaneous
down-conversion in our earlier work \cite{systematic}. This is the
evidence that quantum correlations do not disappear when the
emission is stimulated. In fact, the down-converted photons are
still produced in pairs, although in the beam 2 the photons of the
pairs are added to the bright original coherent beam propagating
in the same direction. The factor 2, appearing in Eq.
(\ref{stimul-cross-corr}) can be interpreted as due to the fact
that, for any down-converted photon of a pair propagating along
direction 2, there is also the original photon of the seed that
stimulated the generation of that pair. Formally it emerges from
the time integration of two identical contribution in
(\ref{stim12}), the explicit one and its complex conjugated.

Since we are interested in a relatively large power of incident
light, we can at first consider detectors without internal gain,
and assume that the charge produced in any detection event is
equal to the single electron charge $q$, i.e. $\langle
q_{k}\rangle=q$ and $\langle q_{k}^{2}\rangle=q^{2}$. Therefore,
according to Eq.s (\ref{stimul-self-corr}) and
(\ref{stimul-cross-corr}), the quantum efficiency can be evaluated
as

\begin{equation}\label{eta-stim}
\eta_{2}=\frac{1}{2}\frac{\langle \delta i_{1}(t)\delta
i_{2}(t+\tau)\rangle}{\langle \delta i_{1}(t) \delta
  i_{1}(t+\tau)\rangle}.
\end{equation}
For detectors in which electrons multiplication occurs the
statistic fluctuations of this process do not allow to use Eq.
(\ref{eta-stim}) for  absolute calibration.  However, it can be
performed by integrating Eq. (\ref{stimul-cross-corr}) over time
$\tau$ \cite{systematic}. It corresponds to evaluating the power
spectrum of the fluctuations at frequencies around zero, namely
much smaller than $1/\tau_{p}$. We would like to stress that in
this case, the assumption $f_{1}(t)=f_{2}(t)=f(t)$ is not
necessary. Since $\int d\tau\mathcal{F}(\tau)=1$ we obtain
\begin{equation}\label{eta-stim2}
\eta_{2}\langle q_{2}\rangle=\frac{1}{2}\frac{\int d\tau\langle
\delta i_{1}(t)\delta i_{2}(t+\tau)\rangle}{\langle
i_{1}\rangle}\:,
\end{equation}
In Eq. (\ref{eta-stim2}) the statistic of the electron gain does
not play any role.

\section{Conclusion}

With the purpose to eliminate some problems that appear when
working with Spontaneous PDC \cite{systematic}, we have extended
our method to higher flux regimes, taking advantage of the
stimulated PDC as a source of bright correlated beams. We show
that this scheme allows one to overcome these former problem: an
important result in view of metrology applications. Indeed this
result allows to link the photon counting regime with the photon
rates of the traditional optical metrology supplying a single
radiometric standard for all these regimes. Finally, we would like
to stress once more that equations \ref{eta-stim} and
\ref{eta-stim2}, between quantum efficiency and current
fluctuations, even if identical to the ones obtained in our
previous work \cite{systematic}, are here derived for a more
general condition of stimulated emission.

\section*{Acknowledgments}

The Turin group acknowledges  the support of  Regione Piemonte
(ricerca scientifica applicata E14) and San Paolo Foundation. The
Russian group acknowledges the support of Russian Foundation for
Basic Research (grant \# 06-02-16393). Both acknowledge the joint
project of Associazione Sviluppo del Piemonte by Grant
RFBR-PIEDMONT \# 07-02-91581-APS.


\begin{thebibliography}{99}
\bibitem{footnote3}The condition is fulfilled if the electronics is the same for the
two detector and has a pass band much smaller than the band of the detectors.
\end{thebibliography}
\end{document}